\documentclass[12pt,english,a4paper]{article}
\usepackage[T1]{fontenc}
\usepackage[latin9]{inputenc}
\usepackage{amsmath}
\usepackage{graphicx}

\makeatletter

\usepackage{float}

\usepackage{epsfig}

\makeatother

\usepackage{babel}

\begin{document}

\makeatother

\title{\textbf{\large Cutoff Regularization Method in Gauge Theories}}

\author{G. Cynolter%
\thanks{E-mail address: cyn@general.elte.hu %
} and E. Lendvai}

\date{MTA-ELTE Theoretical Physics Research Group, E\"otv\"os University, Budapest,
1117 P\'azm\'any P\'eter s\'et\'any 1/A, Hungary }
\maketitle
\begin{abstract}
In quantum field theories divergences generally turn up in loop
calculations. Renormalization is a part of the theory which can be
performed only with a proper regularization. In low energy effective
theories there is a natural cutoff with well defined physical meaning,
but the naive cutoff regularization is unsatisfactory. A Lorentz and gauge symmetry 
preserving regularization method is discussed
in four dimension based on momentum cutoff. First we give an overview
of various regularization methods then the new regularization is introduced.
We use the conditions of gauge invariance or equivalently the freedom
of shift of the loop momentum to define the evaluation of the terms
carrying even number of Lorentz indices, e.g. proportional to $k_{\mu}k_{\nu}$.
The remaining scalar integrals are calculated with a four dimensional
momentum cutoff. The finite terms (independent of the cutoff) are
free of ambiguities coming from subtractions in non-trivial cases.
Finite parts of the result are equal with the results of dimensional
regularization. The proposed method can be applied to various physical
processes where the use of dimensional regularization is subtle or
a physical cutoff is present. As a famous example it is shown that
the triangle anomaly can be calculated unambiguously with this new
improved cutoff. The anticommutator of $\gamma^{5}$ and $\gamma^{\mu}$
multiplied by five $\gamma$ matrices is proportional to terms that
do not vanish under a divergent loop-momentum integral, but cancel
otherwise.
\end{abstract}

\section{Introduction}

Several regularization methods are known and used in quantum field
theory: three and four dimensional momentum cutoff, Pauli-Villars
type, dimensional regularization, lattice regularization, Schwinger's
proper time method and others directly linked to renormalization like
differential renormalization. Dimensional regularization (DREG) \cite{dreg}
is the most popular and most appreciated as it respects the gauge
and Lorentz symmetries. However DREG is not useful in all cases, for
example it is not directly applicable to supersymmetric gauge theories
as it modifies the number of bosons and fermions differently. DREG
gets rid of (does not identify) naive quadratic divergences, which
may be important in low energy effective theories or in the Wilson's
renormalization group method. Another shortcoming is that together
with (modified) minimal subtraction DREG is a {}``mass independent''
scheme, particle thresholds and decoupling are put in the theory by
hand \cite{georgi}. The choice of the ultraviolet regulator always
depends on the problem.

In low energy effective field theories there is an explicit cutoff,
with well defined physical meaning. The cutoff gives the range of
the validity of the model. There are a few implementations in four
dimensional theories: sharp momentum cutoff in 3 or 4 dimensions,
modified operator regularization (based on Schwinger proper time method
\cite{schwinger}). In the Nambu-Jona-Lasinio model different regularizations
proved to be useful calculating different physical quantities \cite{klev}. 

Using a naive momentum cutoff the symmetries are badly violated. The
calculation of the QED vacuum polarization function ($\Pi_{\mu\nu}(q)$)
shows the problems. The Ward identity tells us that $q^{\mu}\Pi_{\mu\nu}(q)=0$,
e.g. in\begin{equation}
\Pi_{\mu\nu}(q)=q_{\mu}q_{\nu}\Pi_{L}(q^{2})-g_{\mu\nu}q^{2}\Pi_{T}(q^{2})\label{eq: Pidef}\end{equation}
 the two coefficients must be the same $\Pi(q^{2})$. Usually the
condition $\Pi(0)=0$ is required to define a subtraction to keep
the photon massless at 1-loop. However this condition is ambiguous
when one calculates at $q^{2}\neq0$ in QED or in more general models.
For example in the case of two different masses in the loop, it just
fixes $\Pi(q^{2},\ m_{1},\ m_{2})$ in the limit of degenerate masses
at $q^{2}=0$. Ad hoc subtractions does not necessarily give satisfactory
results.

There were several proposals how to define symmetry preserving cutoff
regularization. Usual way is to start with a regularization that respects
symmetries and find the connection with momentum cutoff. In case of
dimensional regularization already Veltman observed \cite{veltman}
that the naive quadratic divergences can be identified with the poles
in two dimensions (d=2) besides the usual logarithmic singularities
in d=4. This idea turned out to be fruitful. Hagiwara et al. \cite{hagiwara}
calculated electroweak radiative corrections originating from effective
dimension-six operators, and later Harada and Yamawaki performed the
Wilsonian renormalization group inspired matching of effective hadronic
field theories \cite{harada}. Based on Schwinger's proper time approach
Oleszczuk proposed the operator regularization method \cite{Olesz},
and showed that it can be formulated as a smooth momentum cutoff respecting
gauge symmetries \cite{Olesz,liao}. A momentum cutoff is defined
in the proper time approach in \cite{varin} with the identification
under loop integrals \begin{equation}
k_{\mu}k_{\nu}\rightarrow\frac{1}{d}g_{\mu\nu}k^{2}\label{eq: perdim}\end{equation}
instead%
\footnote{In what follows we denote the metric tensor by $g_{\mu\nu}$ both
in Minkowski and Euclidean space.%
} of the standard $d=4$. The degree of the divergence determines $d$
in the result: $\Lambda^{2}$ goes with $d=2$ and $\ln(\Lambda^{2})$
with $d=4$. This way the authors get correctly the\textit{ divergent
parts,} they checked them in the QED vacuum polarization function
and in the phenomenological chiral model.

Various authors formulated consistency conditions to maintain gauge
invariance during the evaluation of divergent loop integrals. Finite
\cite{gu} or infinite \cite{horejsi,wu1} number of new regulator
terms added to the propagators a'la Pauli-Villars, the integrals are
tamed to have at most logarithmic singularities and become tractable.
Pauli-Villars regularization technique were applied with subtractions
to gauge invariant and chiral models \cite{osipov,ruiz,bernard,bajc}.
Differential renormalization can be modified to fulfill consistency
conditions automatically, it is called constrained differential renormalization
\cite{cdr}. Another method, later proved to be equivalent with the
previous one \cite{equiv}, is called implicit regularization, a recursive
identity (similar to Taylor expansion) is applied and all the dependence
on the external momentum ($q$) is moved to finite integrals. The
divergent integrals contain only the loop momentum, thus universal
local counter terms can cancel the potentially dangerous symmetry
violating contributions \cite{nemes,nemes2}. A strictly four dimensional
approach to quantum field theory is proposed in \cite{fdr}. They
interpreted ultraviolet divergencies as a natural separation between
physical and non-physical degrees of freedom providing gauge invariant
and cutoff independent loop integrals, it was also applied to non-renormalizable
theories \cite{fdrNR}. Gauge invariant regularization is implemented
in exact renormalization group method providing a cutoff without gauge
fixing in \cite{rosten}. Introducing a multiplicative regulator in
the d-dimensional integral, the integrals are calculable in the original
dimension with the tools of DREG \cite{dillig}.

In this chapter we give a definite method in four dimensions to use
a well defined momentum cutoff. We show that there is a tension between
naive application of Lorentz symmetry and gauge invariance. The core
of the problem is that contraction with $g^{\mu\nu}$ cannot necessarily
be interchanged with the integration in divergent cases. The proper
handling of the $k_{\mu}k_{\nu}$ terms in divergent loop-integrals
solves the problems of momentum cutoff regularizations. Working in
strictly four dimensions we use the conditions of respecting symmetries
to define the integrals with free Lorentz indices. Using our method
loop calculations can be reduced to scalar integrals and those can
be evaluated with a sharp momentum cutoff. We give a simple and well
defined algorithm to have unambiguous finite and infinite terms \cite{uj}
dubbed as improved momentum cutoff regularization. The method was
successfully applied earlier to a non-renormalizable theory \cite{fcmlambda,cynnova}.
The results respect gauge (chiral and other) symmetries and the finite
terms agree with the result of DREG. There were various other proposals
to modify the calculation with momentum cutoff to respect Lorentz
and gauge symmetries \cite{Olesz,liao,gu,wu1,rosten}.

An ideal application of the improved cutoff is the unambiguous calculation
of the triangle anomaly in four dimensions presented in \cite{anom}.
Dimensional regularization  \cite{dreg} respects Lorentz and gauge
symmetries, but as it modifies the number of dimensions (at least
in the loops) it is not directly applicable to chiral theories, such
as the standard model or to supersymmetric theories. Continuation
of $\gamma_{5}$ to dimensions $d\neq4$ goes with a $\gamma_{5}$
not anticommuting with the extra elements of gamma matrices, and it
leads to {}``spurious anomalies'', see \cite{collins,spuri,jegerl,korner},
and references therein. The loop integrals using the novel improved
momentum cutoff regularization are invariant to the shift of the loop
momentum, therefore the usual derivation of the ABJ triangle anomaly
would fail in this case. We extend the method to graphs involving
$\gamma_{5}$, and show that the proper handling of the trace of $\gamma_{5}$
and six gamma matrices provides the correct anomaly, the $\left\{ \gamma_{5},\gamma_{\mu}\right\} $
anticommutator does not vanish in special cases under divergent loop
integrals. 

The rest is organized as follows. In section 2 we present how to define
a momentum cutoff using the method of DREG, then we give the gauge
symmetry preserving conditions emerging during the calculation of
the vacuum polarization amplitude. In section 4 we discuss the condition
of independence of momentum routing in loop diagrams. Section 5 shows
that gauge invariance and freedom of shift in the loop momentum have
the same root. Next we show that the conditions are related to vanishing
surface terms. In section 7 we give a definition of the new regularization
method and in section 8 as an example we present the calculation of
a general vacuum polarization function at 1-loop. In section 9 we
show that the QED Ward-Takahashi identity holds at finite order using
the new method. Section 10 deals with the famous triangle anomaly
and the chapter is closed with conclusions.

\section{Momentum cutoff via dimensional regularization}

DREG is very efficient and popular, because it preserves gauge and
Lorentz symmetries. Performing standard steps the integrals are evaluated
in $d=4-2\epsilon$ dimension. Generally the loop momentum integral
is Wick rotated and with a Feynman parameter ($x$) the denominators
are combined, then the order of $x$ and momentum integrals are changed.
Shifting the loop momentum does not generate surface terms and it
leads to spherically symmetric denominator, terms linear in the momentum
are dropped and \eqref{eq: perdim} is used. Singularities are identified
as $1/\epsilon$ poles, naive power counting shows that these are
the logarithmic divergences of the theory. In DREG quadratic or higher
divergences are set identically to zero. However Veltman noticed \cite{veltman}
that quadratic divergences can be calculated in $d=2-2(\epsilon-1)$
in the limit $\epsilon\rightarrow1$. This observation led to a cutoff
regularization based on DREG.

Carefully calculating the one and two point Passarino-Veltman functions
in DREG and in 4-momentum cutoff the divergences can be matched as
\cite{hagiwara,harada} \begin{eqnarray}
4\pi\mu^{2}\left(\frac{1}{\epsilon-1}+1\right) & = & \Lambda^{2},\label{eq: quad}\\
\frac{1}{\epsilon}-\gamma_{E}+\ln\left(4\pi\mu^{2}\right)+1 & = & \ln\Lambda^{2},\label{eq:log}\end{eqnarray}
where $\mu$ is the mass-scale of dimensional regularization and $\gamma_{E}$
is the Euler-Macheroni constant appearing always together with $1/\epsilon$.
The finite part of a divergent quantity is defined as \begin{equation}
f_{{\rm finite}}=\lim_{\epsilon\rightarrow0}\left[f(\epsilon)-R(0)\left(\frac{1}{\epsilon}-\gamma_{E}+\ln4\pi+1\right)-R(1)\left(\frac{1}{\epsilon-1}+1\right)\right],\label{eq:finite}\end{equation}
where $R(0)$, \, $R(1)$ are the residues of the poles at $\epsilon=0,\,1$
respectively. Note that in the usual $\epsilon\rightarrow0$ limit
the left hand side (LHS) of \eqref{eq: quad} vanishes and no quadratic
divergence appears in the original DREG.

The identifications above define a momentum cutoff calculation based
on the symmetry preserving DREG formulae. This cutoff regularization
is well defined, but still relies on DREG. Let us see the main properties
in the calculation of the vacuum polarization function. In $\Pi_{\mu\nu}$
the quadratic divergence is partly coming from a $k_{\mu}k_{\nu}$
term via $\frac{1}{d}\cdot g_{\mu\nu}k^{2}$, which is evaluated at
$d=2$ instead of the $d=4$ in the naive cutoff calculation. The
$\Lambda^{2}$ terms cancel if and only if this term is evaluated
at $d=2$. This is a warning that the usual \begin{equation}
k_{\mu}k_{\nu}\rightarrow\frac{1}{4}g_{\mu\nu}k^{2}\label{eq:negyed}\end{equation}
 substitution during the naive cutoff calculation of divergent integrals
might be too naive, especially as an intermediate step, the Wick rotation
is legal only for finite integrals. A further finite term additional
to the logarithmic singularity is coming from the well known expansion
in $\frac{1}{4-2\epsilon}\frac{1}{\epsilon}\simeq\frac{1}{4}\left(\frac{1}{\epsilon}+\frac{1}{2}\right)$,
and it is essential to retain gauge invariance. We stress that the
shift of the loop momentum is allowed in DREG, an improved cutoff
regularization should inherit it. In the next sections we derive consistency
conditions for general regularizations.

\section{Consistency conditions - gauge invariance}

Calculation in a gauge theory ought to preserve gauge symmetries.
Consider the QED vacuum polarization function with massive electrons.
We start generally (see Fig. 1.) with two fermions with different
masses in the loop \cite{fcmlambda} and restrict it to QED later,

\begin{figure}
\begin{centering}
\includegraphics[scale=0.5]{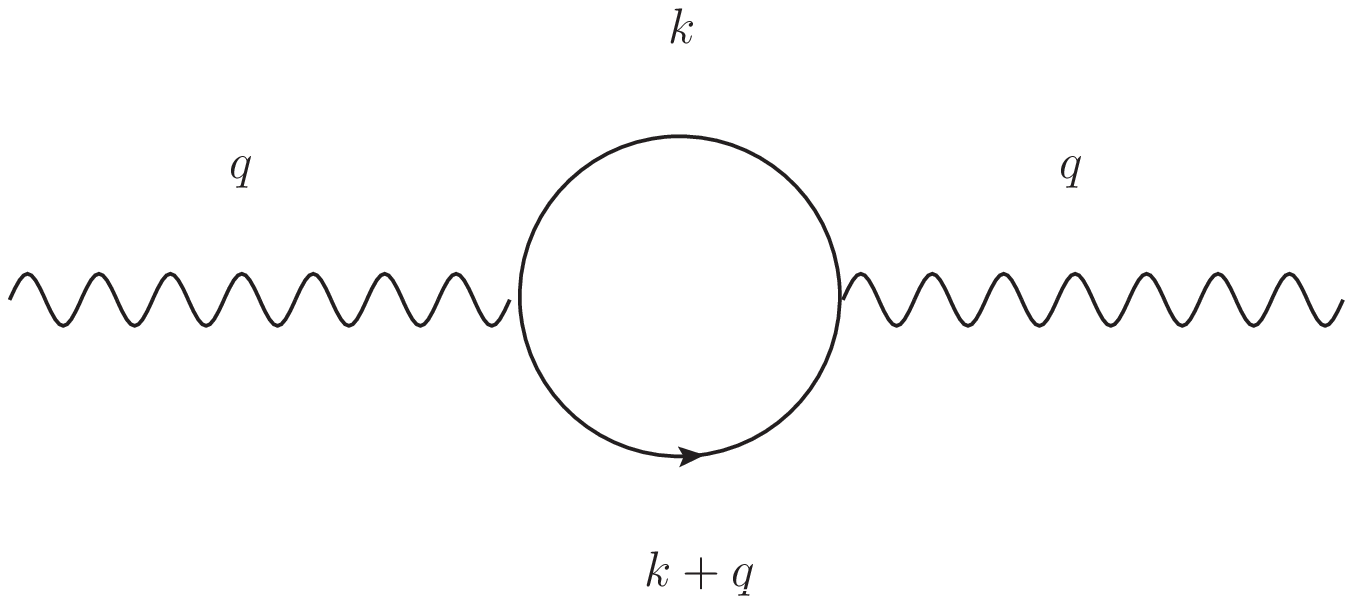}
\par\end{centering}

\begin{centering}
Fig. 1. 1-loop vacuum polarization diagram
\par\end{centering}

\end{figure}
\begin{equation}
i\Pi_{\mu\nu}(q)=-(-ig)^{2}\int\frac{d^{4}k}{(2\pi)^{4}}\hbox{Tr}\left(\gamma_{\mu}\frac{\not k+m_{a}}{k^{2}-m_{a}^{2}}\gamma_{\nu}\frac{\not\not k+\not q+m_{b}}{(k+q)^{2}-m_{b}^{2}}\right).\label{eq:pi1}\end{equation}
 $\Pi_{\mu\nu}$ is calculated with the standard technique, only the
$k_{\mu}k_{\nu}$ terms are considered with care. After performing
the trace, Wick rotating and introducing the Feynman x-parameter the
loop momentum is shifted $(k_{E\mu}+xq_{E\mu})\rightarrow l_{E\mu}$,

\begin{equation}
\Pi_{\mu\nu}=g^{2}\int_{0}^{1}dx\int\frac{d^{4}l_{E}}{(2\pi)^{4}}\frac{2l_{E\mu}l_{E\nu}-g_{\mu\nu}\left(l_{E}^{2}+\Delta\right)-2x(1-x)q_{E\mu}q_{E\nu}+2x(1-x)g_{\mu\nu}q_{E}^{2}}{\left(l_{E}^{2}+\Delta\right)^{2}},\label{eq:pi2}\end{equation}
where $\Delta=x(1-x)q_{E}^{2}+(1-x)m_{a}^{2}+xm_{b}^{2}$. In QED
$m_{a}=m_{b}=m$ and $g=e$, it simplifies to $\Delta_{1}=x(1-x)q_{E}^{2}+m^{2}$.
Having a symmetric denominator and symmetric volume of integration
the terms linear in $l_{E\mu}$ are dropped. After changing the order
of momentum- and x-integration the loop momentum is shifted with x-dependent
values, $xq_{E\mu}$ and sum up the results during the integration.
Different shifts sums up to a meaningful result only if the shift
does not modify the value of the momentum integral (it will be discussed
in the next section). 

In QED the Ward identity tells us, that

\begin{equation}
q^{\mu}\Pi_{\mu\nu}(q)=0.\label{eq:ward}\end{equation}
In \eqref{eq:pi2} the terms proportional to $q_{E}$ fulfill the
Ward-identity \eqref{eq:ward} and what remains is the condition of
gauge invariance\begin{equation}
\int_{0}^{1}dx\int\frac{d^{4}l_{E}}{(2\pi)^{4}}\frac{l_{E\mu}l_{E\nu}}{\left(l_{E}^{2}+\Delta_{1}\right)^{2}}=\frac{1}{2}g_{\mu\nu}\int_{0}^{1}dx\int\frac{d^{4}l_{E}}{(2\pi)^{4}}\frac{1}{\left(l_{E}^{2}+\Delta_{1}\right)}.\label{eq:condgauge}\end{equation}
This condition appeared already in \cite{wu1,nemes}. Any gauge invariant
regulator should fulfill \eqref{eq:condgauge}. It holds in dimensional
regularization and in the momentum cutoff based on DREG of Section
2. In \cite{gu,wu1} a similar relation defined the finite or infinite
Pauli-Villars terms to maintain gauge invariance.

So far the $x$ integrals were not performed. Expanding the denominator
in $q^{2}$, the x-integration can be easily done and we arrive at
a condition for gauge invariance at each order of $q^{2}$. At order
$q^{2n}$ we get (omitting the factor $(2\pi)^{4}$) \begin{equation}
\int d^{4}l_{E}\frac{l_{E\mu}l_{E\nu}}{\left(l_{E}^{2}+m^{2}\right)^{n+1}}=\frac{1}{2n}g_{\mu\nu}\int d^{4}l_{E}\frac{1}{\left(l_{E}^{2}+m^{2}\right)^{n}},\ \ \ \ \ n=1,2,...\label{eq:cn}\end{equation}
 The conditions \eqref{eq:cn} are valid for arbitrary $m^{2}$ mass,
so it holds for any function $\Delta$ independent of the loop momentum
in 1-loop two or n-point functions with arbitrary masses in the propagators.
These conditions mean that in any gauge invariant regularization the
two sides of \eqref{eq:cn} should give the same result. We will use
this condition to define the LHS of \eqref{eq:cn} in the new improved
cutoff regularization. This is the novelty of our regularizations
method.

\section{Consistency conditions - momentum routing}

Evaluating any loops in QFT one encounters the problem of momentum
routing. The choice of the internal momenta should not affect the
result of the loop calculation. The simplest example is the 2-point
function. In \eqref{eq:pi1} there is a loop momentum $k$, and the
external momentum $q$ (see Fig. 1.) is put on one line $(k+q,k)$,
but any partition of the external momentum $(k+q+p,\ \ k+p)$ must
be as good as the original. The arbitrary shift of the loop momentum
should not change the physics. This independence of the choice of
the internal momentum gives a conditions. We will impose it on a very
simple loop integral \begin{equation}
\int d^{4}k\frac{k_{\mu}}{k^{2}-m^{2}}-\int d^{4}k\frac{k_{\mu}+p_{\mu}}{\left(k+p\right)^{2}-m^{2}}=0\label{eq:shift1}\end{equation}
which turns up during the calculation of the 2-point function. Expanding
\eqref{eq:shift1} in powers of $p$ we get a series of condition,
meaningful at $p,\ p^{3},\ p^{5}\ ...$. At linear order we arrive
at \begin{equation}
\int d^{4}k\left(\frac{p_{\mu}}{k^{2}-m^{2}}-2\frac{k_{\mu}k\cdot p}{\left(k^{2}-m^{2}\right)^{2}}\right)=0,\label{eq:c1}\end{equation}
which is equivalent to \eqref{eq:cn} for $n=1$. At order $p^{3}$
a linear combination of two conditions should vanish \begin{equation}
p_{\rho}p_{\alpha}p_{\beta}\int d^{4}k\left[\left(\frac{4k_{\alpha}k_{\beta}}{\left(k^{2}-m^{2}\right)^{3}}-\frac{g_{\alpha\beta}}{\left(k^{2}-m^{2}\right)^{2}}\right)g_{\mu\rho}-4k_{\mu}\left(\frac{2k_{\alpha}k_{\beta}k_{\rho}}{\left(k^{2}-m^{2}\right)^{4}}-\frac{g_{\alpha\beta}k_{\rho}}{\left(k^{2}-m^{2}\right)^{3}}\right)\right]=0.\label{eq:c3}\end{equation}
These two conditions get separated if the freedom of the shift of
the loop momentum is considered in $\int d^{4}k\frac{k_{\mu}}{\left(k^{2}-m^{2}\right)^{2}}$.
At leading order it provides \begin{equation}
p_{\nu}\int d^{4}k\left(\frac{g_{\mu\nu}}{\left(k^{2}-m^{2}\right)^{2}}-4\frac{k_{\mu}k_{\nu}}{\left(k^{2}-m^{2}\right)^{3}}\right)=0,\label{eq:cp}\end{equation}
equivalent with \eqref{eq:cn} for $n=2$. Using \eqref{eq:cp} twice
the second part of the condition \eqref{eq:c3} connects 4 loop momenta
numerators to 2 $k$'s. Symmetrizing the indices we get\begin{equation}
\int d^{4}k\frac{k_{\alpha}k_{\beta}k_{\mu}k_{\rho}}{\left(k^{2}-m^{2}\right)^{4}}=\frac{1}{24}\int d^{4}k\frac{g_{\alpha\beta}g_{\mu\rho}+g_{\alpha\mu}g_{\beta\rho}+g_{\alpha\rho}g_{\beta\mu}}{\left(k^{2}-m^{2}\right)^{2}}.\label{eq:c4}\end{equation}
Invariance of momentum routing provides conditions for symmetry preserving
regularization and these conditions are equivalent with the conditions
coming from gauge invariance.

\section{Gauge invariance and loop momentum shift}

We show at one loop level that gauge invariance of the vacuum polarization
function is equivalent to invariance of a special loop integrand against
shifting the loop momentum \eqref{eq:shift1}. Consider $\Pi_{\mu\nu}$
defined in \eqref{eq:pi1}, performing the trace we get\begin{equation}
i\Pi_{\mu\nu}(q)=-g{}^{2}\int\frac{d^{4}k}{(2\pi)^{4}}\frac{k_{\mu}\left(k_{\nu}+q_{\nu}\right)+k_{\nu}\left(k_{\mu}+q_{\mu}\right)-g_{\mu\nu}\left(k^{2}+k\cdot q-m_{a}m_{b}\right)}{\left(k^{2}-m_{a}^{2}\right)\left((k+q)^{2}-m_{b}^{2}\right)}.\label{eq:pi3}\end{equation}
Specially in QED $m_{a}=m_{b}=m$, gauge invariance requires \eqref{eq:ward},
which simplifies to\begin{equation}
iq^{\nu}\Pi_{\mu\nu}(q)=g^{2}\int\frac{d^{4}k}{(2\pi)^{4}}\left(\frac{k_{\mu}+q_{\mu}}{\left((k+q)^{2}-m^{2}\right)}-\frac{k_{\mu}}{\left(k^{2}-m^{2}\right)}\right)=0.\label{eq:pi4}\end{equation}
This example shows that the Ward identity is fulfilled only if the
shift of the loop momentum does not change the value of the integral,
like in \eqref{eq:shift1}.

In \cite{nemes2} based on the general diagrammatic proof of gauge
invariance it is shown that the Ward identity is fulfilled if the
difference of a general n-point loop and its shifted version vanishes
\begin{equation}
-i\int d^{4}p_{1}Tr\left[\frac{i}{\not p_{n}-m}\gamma^{\mu_{n}}...\frac{i}{\not p_{1}-m}\gamma^{\mu_{1}}-\frac{i}{\not p_{n}+\not q-m}\gamma^{\mu_{n}}...\frac{i}{\not p_{1}+\not q-m}\gamma^{\mu_{1}}\right]=0.\label{eq:ward-shift}\end{equation}
We interpret \eqref{eq:pi4} and \eqref{eq:ward-shift} as a necessary
condition for gauge invariant regularizations.

\section{Consistency conditions - vanishing surface terms}

All the previous conditions are related to the volume integral of
a total derivative\begin{equation}
\int d^{4}k\frac{\partial}{\partial k^{\nu}}\left(\frac{k_{\mu}}{\left(k^{2}+m^{2}\right)^{n}}\right)=\int d^{4}k\left(\frac{k_{\mu}k_{\nu}}{\left(k^{2}+m^{2}\right)^{n+1}}-\frac{1}{2n}g_{\mu\nu}\frac{1}{\left(k^{2}+m^{2}\right)^{n}}\right),\ \ \ n=1,2,...\label{eq:surface}\end{equation}
The total derivative on the LHS leads to surface terms \cite{brigi},
which vanish for finite valued integrals and should vanish for symmetry
preserving regularization. In our improved regularization this will
follow from new definitions. The LHS is in connection with an infinitesimal
shift of the loop momentum $k$, it should be zero if the integral
of the term in the delimiter is invariant against the shift of the
loop momentum. The vanishing of this surface terms reproduces on the
RHS the previous conditions \eqref{eq:c1} and \eqref{eq:cn}. In
\eqref{eq:surface} starting with any odd number of $k$'s in the
numerator we end up with some conditions, three $k$'s for $n=3$
provide \eqref{eq:c4} after some algebra. Starting with even number
of $k_{\mu}$'s in the numerator on the LHS in \eqref{eq:surface}
we get relations between odd number of $k_{\mu}$'s in the numerators,
which vanish separately.

These surface terms all vanish in DREG and give the basis of DREG
respecting Lorentz and gauge symmetries. Vanishing of the surface
term is inherited to any regularization, like improved momentum cutoff,
if the identification \eqref{eq:condgauge} is understood to evaluate
integrals involving even number of free Lorentz indices, e.g. numerators
alike $k_{\mu}k_{\nu}$. The value of integrals with odd number of
$k$'s in the numerator are similarly dictated by symmetry, these
are required to vanish by the symmetry of the integration volume.

\section{Improved momentum cutoff regularization}

We propose a new symmetry preserving regularization based on 4-dimensional
momentum cutoff. During this improved momentum cutoff regularization
method a simple sharp momentum cutoff is introduced to calculate the
divergent scalar integrals in the end. The evaluation of loop-integrals
starts with the usual Wick rotation, Feynman parametrization and loop
momentum shift. The only crucial modification is that the potentially
symmetry violating loop integrals containing explicitly the loop momenta
with free Lorentz indices are calculated with the identification

\begin{equation}
\int d^{4}l_{E}\frac{l_{E\mu}l_{E\nu}}{\left(l_{E}^{2}+\Delta\right)^{n+1}}\rightarrow\frac{1}{2n}g_{\mu\nu}\int d^{4}l_{E}\frac{1}{\left(l_{E}^{2}+\Delta\right)^{n}}\label{eq:rule1}\end{equation}
under the loop integrals or with more momenta using the condition
\eqref{eq:c4} or generalizations of it, like \begin{equation}
\int d^{4}l_{E}\frac{l_{E\mu}l_{E\nu}l_{E\rho}l_{E\sigma}}{\left(l_{E}^{2}+\Delta\right)^{n+1}}\rightarrow\frac{g_{\mu\nu}g_{\rho\sigma}+g_{\mu\rho}g_{\nu\sigma}+g_{\mu\sigma}g_{\nu\rho}}{4n(n-1)}\cdot\int d^{4}l_{E}\frac{1}{\left(l_{E}^{2}+\Delta\right)^{n-1}}.\label{eq:rule2}\end{equation}
The momentum integrals containing further the loop momentum with indices
summed up (e.g. $l_{E}^{2}$) in the numerator are simplified in a
standard way cancelling a factor in the denominator\begin{equation}
\int d^{4}l_{E}\frac{l_{E}^{2}\: l_{E\mu}l_{E\nu}\ldots}{\left(l_{E}^{2}+\Delta\right)^{n+1}}=\int d^{4}l_{E}\frac{l_{E\mu}l_{E\nu}\ldots}{\left(l_{E}^{2}+\Delta\right)^{n}}-\int d^{4}l_{E}\frac{\Delta\: l_{E\mu}l_{E\nu}\ldots}{\left(l_{E}^{2}+\Delta\right)^{n+1}}.\label{eq:rule0}\end{equation}
Integrals with odd number of the loop momenta vanish identically.
These identifications guarantee gauge invariance and freedom of shift
in the loop momentum. Under any regularized momentum integrals the
identifications \eqref{eq:rule1} or generalizations like \eqref{eq:rule2}
are understood as a part of the regularization procedure for $n=1,2,..$.
For finite integrals (non divergent, for high enough $n$) the standard
calculation automatically fulfills (\ref{eq:rule1},\ref{eq:rule2}).
The connection with the standard substitution of free indices is discussed
in Appendix A.

Fulfilling the condition \eqref{eq:cn} via the substitution \eqref{eq:rule1}
the results of momentum cutoff based on DREG of section 2 are completely
reproduced performing the calculation in the physical dimensions $d=4$
\cite{fcmlambda,fcmew}. The next three examples show that the new regularization
provides a robust framework for calculating loop integrals and respects
symmetries.

\section{Vacuum polarization function}

As an example let us calculate the vacuum polarization function of
Fig. 1. in a general gauge theory with fermion masses $m_{a},\ m_{b}$.
Performing the calculation in 4 dimensions generally the Ward identities
(required by the theory) are restored by ambiguous and ad hoc subtractions.
The finite terms of different calculations do not match each other
in the literature, see \cite{maiani}, papers citing it and \cite{silva}.
For sake of simplicity we consider only vector couplings. Performing
the trace in \eqref{eq:pi1} we get \eqref{eq:pi3}. Now we can introduce
a Feynman x-parameter, shift the loop momentum and get \eqref{eq:pi2}
after dropping the linear terms. Generally we are interested in low
energy observables like the precision electroweak parameters and need
the first few terms in the power series of $\Pi_{\mu\nu}(q)$. Using
the rule \eqref{eq:rule1} for $n=1$ and expanding the denominator
in $q^{2}$ the scalar loop and x-integrals can be easily calculated
with a 4-dimensional momentum cutoff ($\Lambda$). The result in this
construction is automatically transverse \begin{equation}
\Pi_{\mu\nu}(q)=\frac{g^{2}}{4\pi^{2}}\left(q^{2}g_{\mu\nu}-q_{\mu}q_{\nu}\right)\left[\Pi(0)+q^{2}\Pi'(0)+...\right].\label{eq:pimunu}\end{equation}
The terms independent of the cutoff completely agree with the results
of DREG \cite{fcmew} the logarithmic singularity can be matched with
the $1/\epsilon$ terms using \eqref{eq:log}. Up to ${\cal O}\left(\frac{m^{2}}{\Lambda^{2}}\right)$
we get \begin{eqnarray}
\Pi(0) & \!\!=\!\! & \frac{1}{4}(m_{a}^{2}+m_{b}^{2})-\frac{1}{2}\left(m_{a}-m_{b}\right)^{2}\ln\left(\frac{\Lambda^{2}}{m_{a}m_{b}}\right)-\label{eq:pi0}\\
 &  & \!\!-\!\!\frac{m_{a}^{4}+m_{b}^{4}-2m_{a}m_{b}\left(m_{a}^{2}+m_{b}^{2}\right)}{4\left(m_{a}^{2}-m_{b}^{2}\right)}\ln\left(\frac{m_{b}^{2}}{m_{a}^{2}}\right).\nonumber \end{eqnarray}
 The first derivative is \begin{eqnarray}
\Pi'(0) & \!\!=\!\!\! & -\frac{2}{9}-\frac{4m_{a}^{2}m_{b}^{2}-3m_{a}m_{b}\left(m_{a}^{2}+m_{b}^{2}\right)}{6\left(m_{a}^{2}-m_{b}^{2}\right)^{2}}+\frac{1}{3}\ln\left(\frac{\Lambda^{2}}{m_{a}m_{b}}\right)+\label{eq:piv0}\\
 &  & +\frac{\left(m_{a}^{2}+m_{b}^{2}\right)\left(m_{a}^{4}-4m_{a}^{2}m_{b}^{2}+m_{b}^{4}\right)+6m_{a}^{3}m_{b}^{3}}{6\left(m_{a}^{2}-m_{b}^{2}\right)^{3}}\ln\left(\frac{m_{b}^{2}}{m_{a}^{2}}\right).\nonumber \end{eqnarray}
The photon remains massless in QED, as in the limit, $m_{a}=m_{b}$
we get $\Pi(0)=0$.

The proposed regularization is robust and gives the same result if
the calculation is organized in a different way. Introducing Feynman
parameters and shifting the loop momentum can be avoided if we need
only the first few terms in the Taylor expansion of $q$. For small
$q$ the second denominator in \eqref{eq:pi3} can be Taylor expanded,
for simplicity we give the expanded integrand for equal masses, up
to ${O}(q^{4})$\begin{eqnarray}
\Pi_{\mu\nu}(q) & = & -g{}^{2}\int\frac{d^{4}k_{E}}{(2\pi)^{4}}\left[2k_{\mu}k_{\nu}\left(\frac{1}{\left(k_{E}^{2}+m^{2}\right)^{2}}-\frac{q_{E}^{2}}{\left(k_{E}^{2}+m^{2}\right)^{3}}+\frac{4\left(k_{E}\cdot q_{E}\right)^{2}}{\left(k_{E}^{2}+m^{2}\right)^{4}}\right)\right.\label{eq:piexp}\\
 & -\negmedspace\! & \left.\frac{2\left(k_{E\mu}q_{E\nu}+k_{E\nu}q_{E\mu}\right)k_{E}\cdot q_{E}}{\left(k_{E}^{2}+m^{2}\right)^{3}}-g_{\mu\nu}\left(\frac{1}{\left(k_{E}^{2}+m^{2}\right)^{2}}-\frac{q_{E}^{2}}{\left(k_{E}^{2}+m^{2}\right)^{3}}+\frac{2\left(k_{E}\cdot q_{E}\right)^{2}}{\left(k_{E}^{2}+m^{2}\right)^{4}}\right)\right].\nonumber \end{eqnarray}
 Taking into account that $k_{E}\cdot q_{E}=k_{E\alpha}q_{E\alpha}$,
\eqref{eq:rule1} and \eqref{eq:rule2} can be used and the remaining
scalar integrals can be easily calculated. The result agrees with
\eqref{eq:pi0} and \eqref{eq:piv0} and the finite terms with DREG
if and only if we use the proposed symmetry preserving substitutions.
Applying the naive $k_{E\mu}k_{E\nu}\rightarrow\frac{1}{4}g_{\mu\nu}k_{E}^{2}$
substitution \eqref{eq:negyed} in both approaches the finite terms
will differ from each other and also from the result of DREG. This
is why finite terms differ from each other in \cite{maiani} and \cite{silva}.

The gauge invariance of the improved momentum cutoff regularization
can be checked directly by the well known identity in QED.

\section{The Ward-Takahashi identity}

In this section we show by explicit calculation that the QED Ward-Takahashi
identity is fulfilled for infinite and finite terms using the proposed
regularization at 1-loop. 
The identity reflects the gauge invariance of the underlying theory.

Following the notation of \cite{peskins} it has to be proved that\begin{equation}
\left.\frac{d\Sigma}{dp_{\mu}}\right|_{\not p=m}=-\left.\delta\Gamma^{\mu}(p,p)\phantom{\frac{1}{2}}\right|_{\not p=m}\:,\label{eq:ward2}\end{equation}
\begin{figure}
\begin{centering}
\includegraphics[scale=0.5]{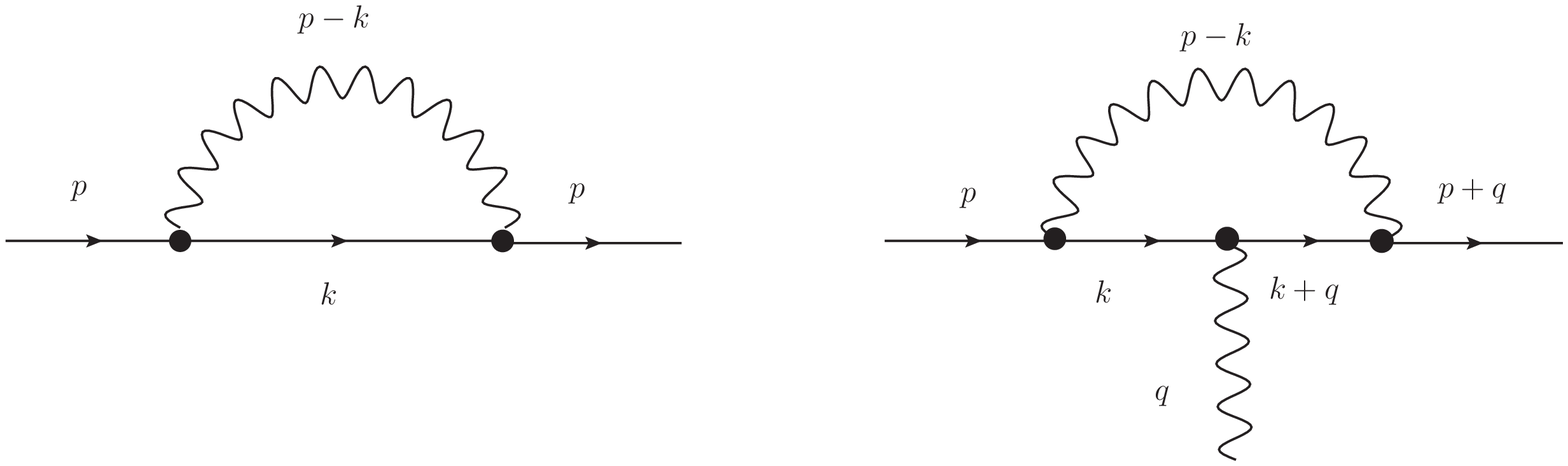}
\par\end{centering}

\centering{}Fig. 2. 1-loop diagrams for the Ward identity
\end{figure}
where $\Sigma$ is the electron self-energy (see Fig. 2. left panel)\begin{equation}
-i\bar{u}(p)\Sigma u(p)=-e^{2}\int_{0}^{1}dx\int\frac{d^{4}l}{(2\pi)^{4}}\bar{u}(p)\frac{-2x\not p+4m}{\left(l^{2}-\Delta_{2}+i\epsilon\right)^{2}}u(p)\:,\label{eq:sigma}\end{equation}
here $\Delta_{2}=-x(1-x)p^{2}+(1-x)m^{2}+x\mu^{2}$, $l=k-xp,$ $m$
is the mass of the electron and $\mu$ is the infrared regulator.\begin{equation}
\left.\frac{d\Sigma}{dp_{\mu}}\right|_{\not p=m}=\frac{\alpha}{2\pi}\int_{0}^{1}dx\left[-x\gamma^{\mu}\left(\ln\left(\frac{\Lambda^{2}}{(1-x)^{2}m^{2}+x\mu^{2}}\right)-1+\frac{2(2-x)(1-x)}{(1-x)^{2}m^{2}+x\mu^{2}}\right)\right]\:,\label{eq:ward2a}\end{equation}
 $\delta\Gamma^{\mu}$ is the electron vertex correction (see Fig.
2. right panel)\begin{equation}
\bar{u}(p')\delta\Gamma^{\mu}u(p)=2ie^{2}\!\int\!\frac{d^{4}k}{(2\pi)^{4}}\frac{\bar{u}(p')\left[\not k\gamma^{\mu}\!(\not k+\not q)+m^{2}\gamma^{\mu}-2m(k+(k\!+\! q))^{\mu}\right]u(p)}{\left((k-p)^{2}+i\epsilon\right)\left((k+q)^{2}-m^{2}+i\epsilon\right)\left(k{}^{2}-m^{2}+i\epsilon\right)}.\label{eq:gamma1}\end{equation}
After using the Dirac equation in the limit $p=p'$ and $q=0$ we
get\begin{eqnarray}
-i\bar{u}(p)\delta\Gamma^{\mu}u(p) & = & 2e^{2}\int_{0}^{1}dxdydz\delta(x+y+z-1)\int\frac{d^{4}l}{(2\pi)^{4}}\nonumber \\
 &  & \times\frac{\bar{u}(p)\left[\not l\gamma^{\mu}\!\not l+(z^{2}-4z+1)m^{2}\gamma^{\mu}\right]u(p)}{\left(l^{2}-\Delta_{3}+i\epsilon\right)^{3}},\label{eq:gamma2}\end{eqnarray}
where $\Delta_{3}=(1-z^{2})m^{2}+z\mu^{2}$ and $l=k-zp$. Here $\not l\gamma^{\mu}\!\!\not l=2l^{\mu}l^{\nu}\gamma_{\nu}-\gamma^{\mu}l^{2}$,
for the first term \eqref{eq:rule1} should be used for $n=2$ or
directly \eqref{eq:i5} from Appendix B. After the momentum and $x,\, y$
integration\begin{equation}
\left.\delta\Gamma^{\mu}\right|_{\not p=m}=\frac{\alpha}{2\pi}\int_{0}^{1}dz\left[(1-z)\left(\ln\left(\frac{\Lambda^{2}}{(1-z)^{2}m^{2}+z\mu^{2}}\right)-1+\frac{(1-4z+z^{2})}{(1-z)m^{2}+z\mu^{2}}\right)\right]\:.\label{eq:gamma3}\end{equation}
The result of the new method is the constant $-1$ after the log,
with the naive calculation using \eqref{eq: standard} one would get
$-1/2$. Calculating the Feynman-parameter integral taking care of
the infrared regulator the identity \eqref{eq:ward2} holds up to
$m^{2}/\Lambda^{2}$ terms at 1-loop \begin{equation}
-\left.\frac{d\Sigma}{dp_{\mu}}\right|_{\not p=m}=\left.\delta\Gamma^{\mu}(p,p)\phantom{\frac{1}{2}}\right|_{\not p=m}=\frac{\alpha}{2\pi}\left(\frac{1}{2}\ln\left(\frac{\Lambda^{2}}{m^{2}}\right)+\ln\left(\frac{\mu^{2}}{m^{2}}\right)+2\right)+{\cal O}\left(\frac{m^{2}}{\Lambda^{2}},\frac{\mu^{2}}{\Lambda^{2}}\right)\:.\label{eq:ward2b}\end{equation}
We have seen that the proposed method provides regularized 1-loop
electron self-energy and vertex correction in QED which fulfill the
Ward-Takakashi identity.

\section{Triangle anomaly}

\begin{figure}
\begin{centering}
\includegraphics[width=11cm]{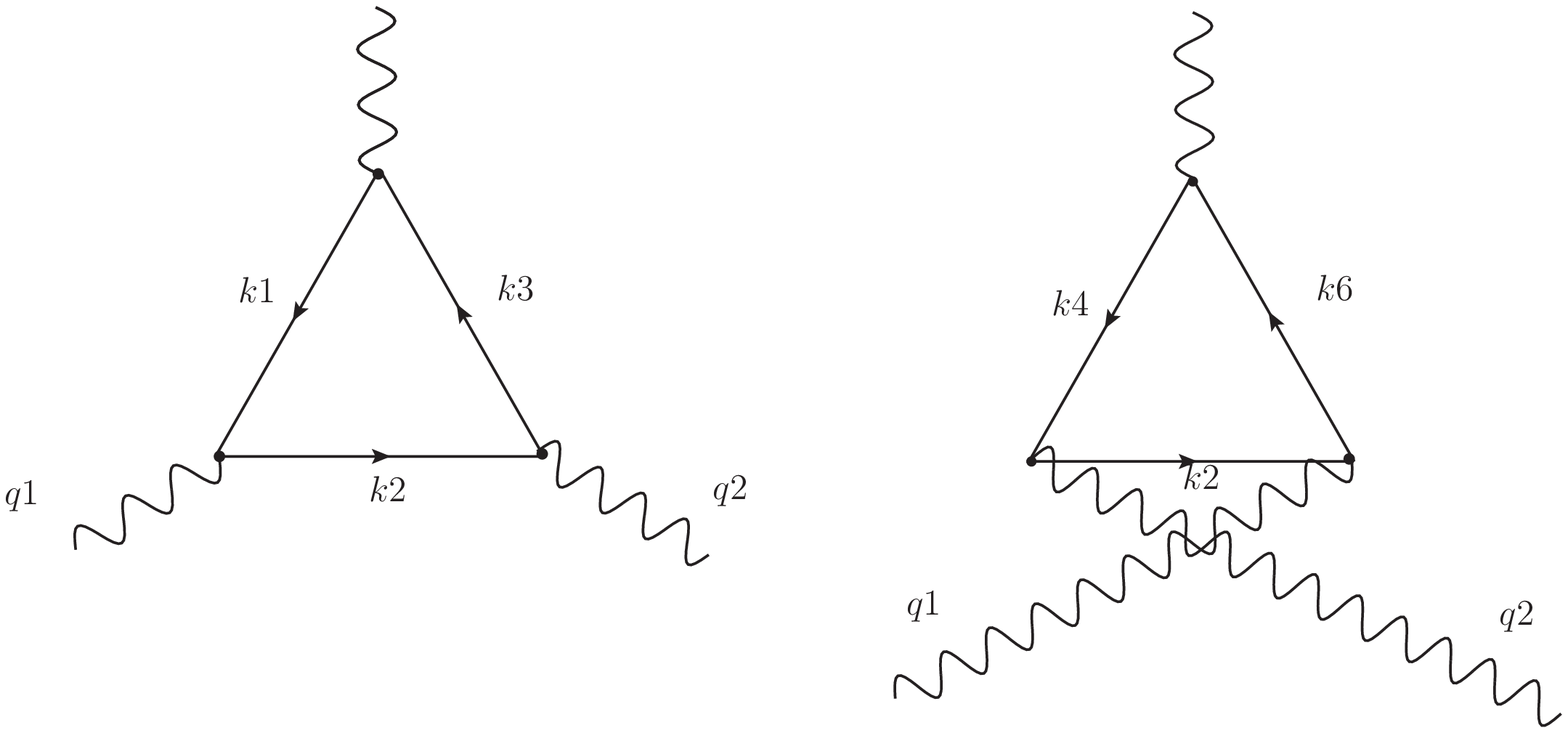}
\par\end{centering}

\centering{}Fig. 3. Feynman graphs contributing to the triangle anomaly,
$k_{2}=k$, $k_{1}=k-q_{1}$, $k_{3}=k+q_{2}$, $k_{4}=k-q_{2}$ and
$k_{6}=k+q_{1}$. 
\end{figure}
In the improved momentum cutoff framework the triangle anomaly has
to be recalculated. The loop integrals using the novel improved
momentum cutoff regularization are invariant to the shift of the loop
momentum, therefore the usual derivation of the ABJ triangle anomaly
would fail in this case (cannot pick up a finite term shifting the
linear divergence). We extend our regularization to graphs involving
$\gamma_{5}$, and show that the proper handling of the trace of $\gamma_{5}$
and six gamma matrices provides the correct anomaly, the $\left\{ \gamma_{5},\gamma_{\mu}\right\} $
anticommutator does not vanish under divergent loop integrals.
In this section we show that the new method provides a well defined result for the famous
triangle anomaly.

Consider the 1-loop triangle graph on the left on Fig. 3.\begin{equation}
T_{1}^{\mu\nu\rho}=e^{2}\int\frac{d^{4}k}{(2\pi)^{4}}Tr\left(\gamma^{5}\frac{\not k-\not q_{1}+m}{\left(k-q_{1}\right)^{2}-m^{2}}\gamma^{\mu}\frac{\not k+m}{k^{2}-m^{2}}\gamma^{\nu}\frac{\not k+\not q_{2}+m}{\left(k+q_{2}\right)^{2}-m^{2}}\gamma^{\rho}\right).\phantom{qq}\label{eq:triang1}\end{equation}
The amplitude of the crossed graph $T_{2}^{\mu\nu\rho}$ is similar
with $(q_{1},\,\mu)$ and $(q_{2},\,\nu)$ interchanged ($T^{\mu\nu\rho}=T_{1}^{\mu\nu\rho}+T_{2}^{\mu\nu\rho}$).
The Ward identities require\begin{eqnarray}
q_{1\mu}T^{\mu\nu\rho} & = & 0,\label{eq:a1}\\
q_{2\nu}T^{\mu\nu\rho} & = & 0,\label{eq:a2}\\
-(q_{1}+q_{2})_{\rho}T^{\mu\nu\rho} & = & 2mT^{5\mu\nu},\label{eq:a3}\end{eqnarray}
where $T^{5\mu\nu}$ corresponds to the same graphs with a pseudoscalar
current instead of the axialvector one. There is a formal proof of
(\ref{eq:a3}). Replace

\begin{equation}
-(q_{1}+q_{2})_{\rho}\gamma^{\rho}\gamma^{5}=-\left(\not k+\not q_{2}-m\right)\gamma^{5}+\left(\not k-\not q_{1}-m\right)\gamma^{5}.\label{eq:alak}\end{equation}
The first term combines with the numerator of the last term in \eqref{eq:triang1}
and cancels the denominator. If \begin{equation}
\left\{ \gamma^{\mu},\gamma^{5}\right\} =0\label{eq:AC}\end{equation}
assumed, then the second term in \eqref{eq:alak} is $-\left(\not k-\not q_{1}-m\right)\gamma^{5}=+\gamma^{5}\left(\not k-\not q_{1}-m\right)+2m\gamma^{5}$.
Here the first term cancels the adjacent fraction in \eqref{eq:triang1}
and the second term gives the right hand side of \eqref{eq:a3}. The
$\left(-(q_{1}+q_{2})_{\rho}T_{1}^{\mu\nu\rho}-2mT_{1}^{5\mu\nu}\right)$
difference is\begin{equation}
e^{2}\int\frac{d^{4}k}{(2\pi)^{4}}\left(\gamma^{5}\frac{\not k-\not q_{1}+m}{\left(k-q_{1}\right)^{2}-m^{2}}\gamma^{\mu}\frac{\not k+m}{k^{2}-m^{2}}\gamma^{\nu}+\gamma^{5}\gamma^{\mu}\frac{\not k+m}{k^{2}-m^{2}}\gamma^{\nu}\frac{\not k+\not q_{2}+m}{\left(k+q_{2}\right)^{2}-m^{2}}\right).\label{eq:diff}\end{equation}
Shifting $k\rightarrow k+q_{1}$ in the first term and passing $\gamma^{\mu}$
through $\gamma^{5}$ (using again \eqref{eq:AC}) to the back of
the second term we arrive to a formula that is totally antisymmetric
under the interchange of $(q_{1},\,\mu)$ and $(q_{2},\,\nu)$, and
thus adding the crossed graph ($T_{2}^{\mu\nu\rho}$) the result vanishes.
Similarly \eqref{eq:a1} and \eqref{eq:a2} can be proven but in this
case \eqref{eq:AC} is not needed to apply, because the terms leading
to cancellation are not separated by a factor of $\gamma^{5}$. The
loop momentum can be shifted, this is a fundamental property of the
improved momentum cutoff regularization.

However (\ref{eq:a1}-\ref{eq:a3}) cannot be all true. Pauli-Villars
regularization or careful simple momentum cutoff calculation identifies
a finite anomaly term when shifting the linearly divergent integral.
There is still a remaining ambiguity in connection with momentum routing
and which identity contains the anomaly term in (\ref{eq:a1}-\ref{eq:a3}).
At the same time in improved momentum cutoff or DREG \eqref{eq:a1}
and \eqref{eq:a2} holds but the proof of \eqref{eq:a3} is false%
\footnote{Functional integral derivation of the anomaly shows that the Ward
identity corresponding to the axial vector current \eqref{eq:a3}
must be anomalous \cite{Fuji}. %
}, it relies additionally on \eqref{eq:AC}. This is the first sign
that the naive anticommutator \eqref{eq:AC} cannot be used in all
situations.

The explicit calculation of the triangle diagram \eqref{eq:triang1}
is based on the evaluation of the trace of $\gamma^{5}$ with six
$\gamma$'s. There are various methods to calculate this trace with
superficially different terms at the end. The different results of
the trace can be transformed to each other using the Schouten identity,
involving two loop momenta it reads\begin{equation}
-k^{2}\epsilon_{\mu\nu\lambda\rho}+k^{\alpha}k_{\mu}\epsilon_{\alpha\nu\lambda\rho}+k_{\nu}k^{\alpha}\epsilon_{\mu\alpha\lambda\rho}+k_{\lambda}k^{\alpha}\epsilon_{\mu\nu\alpha\rho}+k_{\rho}k^{\alpha}\epsilon_{\mu\nu\lambda\alpha}=0.\label{eq:schouten}\end{equation}
In the present method this identity cannot be used for the loop momentum
($k$) of a divergent integral before applying the identifications
\eqref{eq:rule1} or \eqref{eq:rule2}, because it would mix free
Lorentz indices and contracted indices, which must be evaluated in
a different way (DREG faces the same difficulty). After performing
the identifications \eqref{eq:rule1} and \eqref{eq:rule2} the quadratic
loop momenta factors cancel with the denominators. The remaining formula
contains the loop momentum in the numerators at maximum linearly,
the corresponding Schouten identity can be applied. The root of the
problem is that in case of divergent integrals the totally antisymmetric
tensor $\epsilon_{\mu\nu\lambda\rho}$ cannot taken out of the integral,
similarly to the case of $g_{\mu\nu}$ in the previous section. No
such problem emerges for finite integrals. 

The breakdown of the early application of the Schouten identity forces
us to choose one dedicated calculation of the trace. The trace is
calculated not using the anticommutator \eqref{eq:AC}, only \begin{equation}
\left\{ \gamma_{\mu},\gamma_{\nu}\right\} =2g_{\mu\nu}\,,\label{eq:clifford}\end{equation}
and general properties of the trace. The unambiguous result is \begin{eqnarray}
 &  & \frac{1}{4}\mathrm{Tr}\left[\gamma_{5}\gamma_{\alpha}\gamma_{\mu}\gamma_{\beta}\gamma_{\nu}\gamma_{\rho}\gamma_{\lambda}\right]=\epsilon_{\alpha\mu\beta\nu}g_{\rho\lambda}-\epsilon_{\alpha\mu\beta\rho}g_{\nu\lambda}+\epsilon_{\alpha\mu\nu\rho}g_{\beta\lambda}-\epsilon_{\alpha\beta\nu\rho}g_{\mu\lambda+}\nonumber \\
 &  & +\epsilon_{\mu\beta\nu\rho}g_{\alpha\lambda}-\epsilon_{\lambda\alpha\mu\beta}g_{\rho\nu}+\epsilon_{\lambda\alpha\mu\nu}g_{\rho\beta}-\epsilon_{\lambda\alpha\beta\nu}g_{\rho\mu}+\epsilon_{\lambda\mu\beta\nu}g_{\rho\alpha}-\epsilon_{\lambda\rho\alpha\mu}g_{\nu\beta}+\nonumber \\
 &  & +\epsilon_{\lambda\rho\alpha\beta}g_{\nu\mu}-\epsilon_{\lambda\rho\mu\beta}g_{\nu\alpha}+\epsilon_{\lambda\rho\nu\alpha}g_{\mu\beta}-\epsilon_{\lambda\rho\nu\mu}g_{\alpha\beta}+\epsilon_{\lambda\rho\nu\beta}g_{\alpha\mu}.\label{eq:tr6}\end{eqnarray}
It reflects the complete Lorentz structure of the $\gamma$ matrices
in the trace. This choice of the trace also appeared in earlier papers
without detailed argumentations \cite{polon,mawu}. All different
calculations of the trace are in agreement with each other and with
\eqref{eq:tr6} if \eqref{eq:AC} is modified. $\gamma_{5}$ and $\gamma_{\mu}$
does not always anticommute (rather the anticommutator picks up terms
proportional to the sum of a few Schouten identities.) Explicitly,
the following definition will eliminate all the ambiguities burdening
the calculation of the trace of $\gamma_{5}$ and six $\gamma$'s
\begin{eqnarray}
\mathrm{Tr}\left[\left\{ \gamma_{\rho},\gamma_{5}\right\} \gamma_{\lambda}\gamma_{\alpha}\gamma_{\mu}\gamma_{\beta}\gamma_{\nu}\right]=\mathrm{2Tr}\left[g_{\nu\rho}\gamma_{5}\gamma_{\lambda}\gamma_{\alpha}\gamma_{\mu}\gamma_{\beta}-g_{\beta\rho}\gamma_{5}\gamma_{\lambda}\gamma_{\alpha}\gamma_{\mu}\gamma_{\nu}+\right.\nonumber \\
\left.+g_{\mu\rho}\gamma_{5}\gamma_{\lambda}\gamma_{\alpha}\gamma_{\beta}\gamma_{\nu}-g_{\alpha\rho}\gamma_{5}\gamma_{\lambda}\gamma_{\mu}\gamma_{\beta}\gamma_{\nu}+g_{\lambda\rho}\gamma_{5}\gamma_{\alpha}\gamma_{\mu}\gamma_{\beta}\gamma_{\nu}\right].\label{eq:trg5}\end{eqnarray}
The above anticommutator is defined only under the trace. \eqref{eq:trg5}
can be understood as the $\left\{ \gamma_{5},\gamma_{\rho}\right\} $
anticommutator is defined by picking up all the terms when moving
$\gamma_{\rho}$ all the way round through the other five $\gamma$'s.
Evaluating the trace the right hand side is proportional to Schouten
identities. Under a divergent loop integral it will not vanish in
the present method (nor in DREG). The nontrivial anticommutator contributes
to the triangle anomaly but vanishes in non-divergent cases and for
less $\gamma$'s. The amplitude of the triangle diagrams can be calculated
with the definition of the trace \eqref{eq:tr6} and the identifications
\eqref{eq:rule1}, \eqref{eq:rule2}. Finally we arrive at the extra
anomaly term in \eqref{eq:a3}.

In what follows we calculate directly the anomaly term missing in
\eqref{eq:a3}. We use \eqref{eq:alak} and move $\left(\not k-\not q_{1}-m\right)$
from the back to the front in \eqref{eq:triang1} using \eqref{eq:clifford}.
Without this trick the same result is obtained evaluating the trace
of six $\gamma$'s and $\gamma^{5}$ using \eqref{eq:tr6}, which
is consistent with non-anticommuting $\gamma^{5}$ in this special
case, see \eqref{eq:trg5}. \begin{eqnarray}
-(q_{1}+q_{2})_{\rho}T_{1}^{\mu\nu\rho} & = & e^{2}\int\frac{d^{4}k}{(2\pi)^{4}}\mathrm{Tr}\left[\phantom{\gamma^{\mu}\frac{1}{\not k+m}}\right.\nonumber \\
 &  & -\gamma^{5}\frac{1}{\not k-\not q_{1}-m}\gamma^{\mu}\frac{1}{\not k-m}\gamma^{\nu}+\gamma^{5}\gamma^{\mu}\frac{1}{\not k+m}\gamma^{\nu}\frac{1}{\not k+\not q_{2}+m}+\phantom{++}\nonumber \\
 &  & +2\gamma^{5}\frac{1}{\not k-\not q_{1}-m}\gamma^{\mu}\frac{1}{\not k-m}\gamma^{\nu}\frac{(k-q_{1})(k+q_{2})}{(k+q_{2})^{2}-m^{2}}-\nonumber \\
 &  & -2\gamma^{5}\frac{1}{\not k-\not q_{1}-m}\gamma^{\mu}\frac{1}{\not k-m}\gamma^{\nu}\frac{(k^{\nu}-q_{1}^{\nu})}{\not k+\not q_{2}+m}+\nonumber \\
 &  & +2\gamma^{5}\frac{1}{\not k-\not q_{1}-m}\gamma^{\mu}\gamma^{\nu}\frac{1}{\not k+\not q_{2}+m}\frac{(k-q_{1})(k)}{k^{2}-m^{2}}-\nonumber \\
 &  & \left.-2\gamma^{5}\frac{1}{\not k-\not q_{1}-m}\frac{1}{\not k+m}\gamma^{\nu}\frac{(k^{\mu}-q_{1}^{\mu})}{\not k+\not q_{2}+m}\right].\label{eq:15}\end{eqnarray}
With algebraic manipulations using the antisymmetry of the trace including
$\gamma_{5}$ and four $\gamma$'s we can group the terms\begin{eqnarray}
-(q_{1}+q_{2})_{\rho}T_{1}^{\mu\nu\rho} &  & =e^{2}\int\frac{d^{4}k}{(2\pi)^{4}}\mathrm{Tr\gamma^{5}\left[\frac{\not k\not q_{1}\gamma^{\mu}\gamma^{\nu}}{N_{1}N_{2}}+\frac{\not k\not q_{2}\gamma^{\mu}\gamma^{\nu}}{N_{2}N_{3}}+2m^{2}\frac{\not q_{1}\not q_{2}\gamma^{\mu}\gamma^{\nu}}{N_{1}N_{2}N_{3}}+\right.\phantom{11}}\nonumber \\
 &  & +\frac{2}{N_{1}N_{2}N_{3}}\left\{ -\not q_{1}\not q_{2}\gamma^{\mu}\gamma^{\nu}\cdot k^{2}+\not k\not q_{2}\gamma^{\mu}\gamma^{\nu}\cdot kq_{1}+\phantom{\frac{2}{2}}\right.\nonumber \\
 &  & \left.\phantom{\frac{2}{2}}+\not q_{1}\not k\gamma^{\mu}\gamma^{\nu}\cdot kq_{2}+\not q_{1}\not q_{2}\not k\gamma^{\nu}\, k^{\mu}+\not q_{1}\not q_{2}\gamma^{\mu}\not k\, k^{\nu}\right\} +\nonumber \\
 &  & +\frac{2}{N_{1}N_{2}N_{3}}\left(+\not\not k\not q_{1}\gamma^{\mu}\gamma^{\nu}\cdot q_{1}q_{2}-\not q_{2}\not q_{1}\gamma^{\mu}\gamma^{\nu}\cdot kq_{1}-\phantom{\frac{2}{2}}\right.\nonumber \\
 &  & \left.\left.\phantom{\frac{2}{2}}-\not\not k\not q_{2}\gamma^{\mu}\gamma^{\nu}\cdot q_{1}^{2}-\not k\not q_{1}\not q_{2}\gamma^{\nu}\, q_{1}^{\mu}-\not k\not q_{1}\gamma^{\mu}\not q_{2}\, q_{1}^{\nu}\right)\right],\label{eq:long}\end{eqnarray}
where $N_{1}=\left((k-q_{1})^{2}-m^{2}\right)$, $N_{2}=\left(k^{2}-m^{2}\right)$
and $N_{3}=\left((k+q_{2})^{2}-m^{2}\right)$. The first two terms
vanish after performing the trace and the integral (they are proportional
to $\epsilon^{\mu\nu\alpha\beta}q_{1\alpha}q_{1\beta}$ and $\epsilon^{\mu\nu\alpha\beta}q_{2\alpha}q_{2\beta}$
respectively). The third one gives $2m$ times the pseudoscalar amplitude
$T^{5\mu\nu}=T_{1}^{5\mu\nu}+T_{2}^{5\mu\nu}$,\begin{equation}
T_{1}^{5\mu\nu}=-m\epsilon^{\mu\nu\alpha\beta}q_{1\alpha}q_{2\beta}e^{2}\int\frac{d^{4}k}{(2\pi)^{4}}\left[\frac{1}{N_{1}N_{2}N_{3}}\right],\label{eq:pvv1}\end{equation}
we get $T_{2}^{5\mu\nu}$ interchanging $(q_{1},\mu)\leftrightarrow(q_{2},\nu)$
in the integrand.

The last five terms in \eqref{eq:long} contain \textit{one factor}
of the loop momentum ($k$) and after tracing vanish by the Schouten
identity, the loop integration does not spoil the cancellation. The
contribution of the one but last five terms in the curly bracket does
not vanish. It contains \textit{two factor} of the loop momentum,
and it is proportional to the Schouten identity \eqref{eq:schouten}
broken under the divergent loop integral. Calculating it with the
improved momentum cutoff of Section 2 using the formulas of the Appendix
(or with DREG) we get the anomaly term

\begin{equation}
-(q_{1}+q_{2})_{\rho}T^{\mu\nu\rho}=2mT^{5\mu\nu}-i\frac{e^{2}}{2\pi^{2}}\epsilon^{\mu\nu\alpha\beta}q_{1\alpha}q_{2\beta}.\label{eq:a03A}\end{equation}
In the case of the naive substitution \eqref{eq:negyed} the Schouten
identity \eqref{eq:schouten} is satisfied, the curly bracket vanishes.
(In that case with simple momentum cutoff the anomaly term originates
from shifting the linearly divergent first two terms in \eqref{eq:long},
but the result depends on momentum routing.) The presented method
identifies without ambiguity the value of the anomaly in the axial-vector
current and leaves the vector currents anomaly free without any further
assumptions.

\section{Conclusions}

We have presented in this chapter a new method for the reliable calculation
of divergent 1-loop diagrams (even involving $\gamma_{5}$) with four
dimensional momentum cutoff. Various conditions were derived to maintain
gauge symmetry, to have the freedom of momentum routing or shifting
the loop momentum. These conditions were known by several authors
\cite{gu,wu1,nemes,nemes2}. Our new proposal is that these conditions
will be satisfied during the regularization process if terms proportional
to loop momenta with even number of free Lorentz indices (e.g. $\sim k_{\mu}k_{\nu}$)
are calculated according to the special identifications \eqref{eq:rule1}
and \eqref{eq:rule2} or generalizations thereof. In the end the scalar
integrals are calculated with a simple momentum cutoff. The calculation
is robust - at least at 1-loop level - as we have shown via the fermionic
contribution to the vacuum polarization function. The finite terms
agree with the one in dimensional regularization in all examples.
The connection with DREG is more transparent if one uses alternatively
the $k_{\mu}k_{\nu}\rightarrow\frac{1}{d}g_{\mu\nu}k^{2}$ or \eqref{eq:perd2}
substitution and $d$ takes different values determined by the degree
of divergence in each term (\ref{eq:quad2}, \ref{eq:log2}, \ref{eq:fin2}).
We stress that this new regularization stands without DREG as the
substitutions \eqref{eq:rule1}, \eqref{eq:rule2} and scalar integration
with a cutoff are independent of DREG. The success of both regularizations
based on the property that they fulfill the consistency conditions
of gauge invariance and momentum shifting.

At 1-loop the finite terms in the improved momentum cutoff are found
to be equivalent with DREG. For practical calculations we propose
to use the same renormalization scheme, $MS$ or $\overline{MS}$
subtractions plus BPHZ forest formula as with DREG. DREG is not just
the generally used method, but it is proved to be a mathematically
rigorous regularization within the Epstein-Glaser framework \cite{falk}.
The equivalence of the results of the proposed method and DREG gives
a hint that the improved cutoff method with e.g. $MS$ subtraction
and BPHZ can be used as a renormalization scheme for more complicated
diagrams.

Regularization schemes based on consistency conditions have been applied
to more involved cases. Constrained differential renormalization is
useful in supersymmetric \cite{drsusy} and non-Abelian gauge theories,
it fulfills Slavnov-Taylor identities at one and two loops \cite{drym}.
Implicit regularization \cite{nemes,nemes2} requires the same conditions
as we used and it was successfully applied to the Nambu-Jona-Lasinio
model \cite{nemes} and to higher loop calculations in gauge theory.
It was shown that the conditions guarantee gauge invariance generally
and the Ward identities are fulfilled explicitly in QED at two-loop
order \cite{nemes2}. In an effective composite Higgs model, the Fermion
Condensate Model \cite{fcm,cynnova} oblique radiative corrections (S and
T parameters) were calculated in DREG and with the improved cutoff,
too, the finite results completely agree. The calculation involved
vacuum polarization functions with two different fermion masses and
no ambiguity appeared \cite{fcmlambda,fcmew}. 

As an application the triangle anomaly was calculated within the 4
dimensional improved momentum cutoff framework. The property that
the loop-integrals are invariant under the shift of the loop momentum
spoils the usual derivation of the ABJ anomaly in the presence of
a cutoff. We calculated the trace \eqref{eq:tr6} corresponding to
the triangle graphs of Fig. 3. ($\gamma_{5}$ and six $\gamma$'s)
and the Ward identity \eqref{eq:a03A} ($\gamma_{5}$ and four $\gamma$'s)
exploiting only the standard anticommutators of the $\gamma$ matrices
\eqref{eq:clifford} and not using the $\gamma^{\mu},\,\gamma^{5}$
anticommuting relation. It turns out that different evaluations of
the trace agree with each other if and only if $\gamma^{5}$ does
not always anticommute with $\gamma^{\mu}$, rather $\left\{ \gamma^{\mu},\gamma^{5}\right\} $
picks up terms proportional to the Schouten identity \eqref{eq:trg5}
if it is multiplied with five more $\gamma$'s under the trace. The
trace of the $\left\{ \gamma^{\mu},\gamma^{5}\right\} $ anticommutator
multiplied with three $\gamma$'s vanishes, as the definition of $\mathrm{Tr}(\gamma^{5}\gamma_{\alpha}\gamma_{\beta}\gamma_{\mu}\gamma_{\nu})$
is unambiguous. The right hand side of \eqref{eq:trg5} is only non-vanishing
if it is under a divergent loop momentum integral. In four dimensional
field theory the nontrivial properties of $\gamma^{5}$ and $\gamma$'s
appear first time in the divergent triangle diagram. 

Traces involving $\gamma^{5}$ and even number of $\gamma$'s can
be calculated in the same manner avoiding the anticommutation of $\gamma^{\mu}$and
$\gamma^{5}$. First the order of $\gamma^{\nu}$'s are reversed applying
\eqref{eq:clifford} then using the cyclicity of the trace we get
back the original trace in the reversed order, the difference gives
the trace twice. This way the $\left\{ \gamma^{\mu},\gamma^{5}\right\} $
anticommutator can be defined, it will not vanish generally. If it
is multiplied with (2n+1) $\gamma$'s it is equal to the sum of (2n+1)
trace involving $\gamma^{5}$ and (2n) $\gamma$'s, see \eqref{eq:trg5}.
It is well known that the general properties of the trace and $\left\{ \gamma^{\mu},\gamma^{5}\right\} =0$
are in conflict with each other, this led to the 't Hooft-Veltman
scheme \cite{dreg,collins}. Our proposal similarly modifies $\left\{ \gamma^{\mu},\gamma^{5}\right\} $
but works in four dimensions and the modifications come into action
only under divergent loop integrals involving enough number of $\gamma$
matrices. There were attempts to preserve $\left\{ \gamma^{\mu},\gamma^{5}\right\} =0$,
but in that case the cyclicity of the trace was lost \cite{korner}.
We have shown with the new method that the vector currents are conserved
and the axial vector current is anomalous, and no ambiguity appears. 

The new regularization is advantageous in special loop-calculations
where one wants to remain in four dimensions, keep the cutoff of the
model, like in effective theories, in derivation of renormalization
group equations, in extra dimensional scenarios or in models explicitly
depending on the space-time dimensions, like supersymmetric theories.
Similar approaches succeeded in the calculation of the anomalous decay
of the Higgs boson to two photons in four dimensions, where gauge
invariance is crucial \cite{hggfdr,hggkile}. We argue that the method
can be successfully used in higher order calculations containing terms
up to quadratic divergences in (non-Abelian) gauge theories, as it
allows for shifts in the loop momenta, which guarantees the 't Hooft
identity \cite{nemes2,thooft}. This symmetry preserving method can
be used also in automatized calculations (similar to \cite{autom})
as even the Veltman-Passarino functions \cite{vp} can be defined
with the improved cutoff. The strength of the improved momentum cutoff
method is that it can be used in theories with quadratic divergencies
important for example in gauge theories including gravitational interactions
\cite{toms}. The calculation in the Einstein-Maxwell system was presented
in \cite{impgr} and quadratic contributions to the photon 2-point
function were identified but after renormalization they vanished and
did not change the original running of the gauge coupling.

\appendix

\section{Connection with simple momentum cutoff}

What is the relation of the new method with the standard (textbook)
\begin{equation}
k_{\mu}k_{\nu}\rightarrow\frac{1}{4}g_{\mu\nu}k^{2}\label{eq: standard}\end{equation}
 substitution? We have to modify it in case of divergent integrals
to respect gauge symmetry, i.e to fulfill \eqref{eq:cn}. Lorentz
invariance dictates that in \eqref{eq:cn} the LHS must be proportional
to the only available tensor $g_{\mu\nu}$, i.e. \begin{equation}
l_{E\mu}l_{E\nu}\rightarrow\frac{1}{d}g_{\mu\nu}l_{E}^{2}\label{eq:perd1}\end{equation}
 can be used, where $d$ is a number to determine%
\footnote{The usual method is to calculate the trace (and get d=4), but interchanging
the order of tracing (multiplication with $g^{\mu\nu}$) and calculating
the divergent integrals cannot be proven to be valid.%
}. Now both sides of equation \eqref{eq:cn} can be calculated with
simple 4-dimensional momentum cutoff. The different powers of $\Lambda$
can be matched on the two sides, and for $n=1$ we get the following
conditions (from gauge invariance) for the value of $d$,\begin{eqnarray}
\frac{1}{d}\Lambda^{2} & \rightarrow & \frac{1}{2}\Lambda^{2},\label{eq:quad2}\\
\frac{1}{d}\ln\left(\frac{\Lambda^{2}+m^{2}}{m^{2}}\right) & \rightarrow & \frac{1}{4}\left(\ln\left(\frac{\Lambda^{2}+m^{2}}{m^{2}}\right)+\frac{1}{2}\right),\label{eq:log2}\\
\frac{1}{d} & \rightarrow & \frac{1}{4}\ \ \mathrm{for\ finite\ terms}.\label{eq:fin2}\end{eqnarray}

We see that for finite valued integrals when the Wick-rotation is
legal, the condition \eqref{eq:cn} and the rule \eqref{eq:rule1}
gives the usual substitution \eqref{eq: standard}, but for divergent
cases we get back the identification partially found by \cite{hagiwara,harada,varin}
and others. Quadratic divergence goes with $d=2$, logarithmic divergence
goes with $d=4$ plus a finite term (a shift), it is the $+1$ in
equation \eqref{eq:log}. For more than 2 even number of indices generalizations
of \eqref{eq:perd1} should be used, for example in case of 4 indices
the \begin{equation}
l_{E\mu}l_{E\nu}l_{E\rho}l_{E\sigma}\rightarrow\frac{1}{d(d+2)}\cdot\left(g_{\mu\nu}g_{\rho\sigma}+g_{\mu\rho}g_{\nu\sigma}+g_{\mu\sigma}g_{\nu\rho}\right)l_{E}^{4}.\label{eq:perd2}\end{equation}
substitution works.

We emphasize again that for non-divergent integrals the rules \eqref{eq:rule1}
and \eqref{eq:rule2} give the same result as the usual calculation
\eqref{eq: standard}.

\section{Basic integrals}

In this appendix we list the basic divergent integrals calculated
by the regularization proposed in this paper. In the following formulae
$m^{2}$ can be any loop momentum $(k)$ independent expression depending
on the Feynman x parameter, external momenta, etc., e.g. $\Delta(x,q,m_{a},m_{b}).$
The regulated integrals are denoted by $\int_{\Lambda\mathrm{\mathrm{reg}}}$
meaning $\int_{\left|k_{E}\right|\leq\Lambda}$, the integration is
understood for Euclidean momenta with absolute value below $\Lambda$.
The integrals \eqref{eq:i1} and \eqref{eq:i4} are just given for
comparison, those calculated with a simple momentum cutoff. \begin{eqnarray}
\int_{\Lambda\mathrm{\mathrm{reg}}}\frac{d^{4}k}{i(2\pi)^{4}}\frac{1}{k^{2}-m^{2}} & \!\!=\!\! & -\frac{1}{(4\pi)^{2}}\left(\Lambda^{2}-m^{2}\ln\left(\frac{\Lambda^{2}+m^{2}}{m^{2}}\right)\right)\label{eq:i1}\\
\int_{\Lambda\mathrm{\mathrm{reg}}}\frac{d^{4}k}{i(2\pi)^{4}}\frac{k_{\mu}k_{\nu}}{\left(k^{2}-m^{2}\right)^{2}} & \!\!=\!\! & -\frac{1}{(4\pi)^{2}}\frac{g_{\mu\nu}}{2}\left(\Lambda^{2}-m^{2}\ln\left(\frac{\Lambda^{2}+m^{2}}{m^{2}}\right)\right)\label{eq:i2}\\
\int_{\Lambda\mathrm{\mathrm{reg}}}\frac{d^{4}k}{i(2\pi)^{4}}\frac{k_{\mu}k_{\nu}k_{\rho}k_{\sigma}}{\left(k^{2}-m^{2}\right)^{3}} & \!\!=\!\! & \!\!-\frac{1}{(4\pi)^{2}}\frac{g_{\mu\nu}g_{\rho\sigma}\!+\! g_{\mu\rho}g_{\nu\sigma}\!+\! g_{\mu\sigma}g_{\nu\rho}}{8}\left(\!\Lambda^{2}-m^{2}\ln\!\left(\frac{\Lambda^{2}+m^{2}}{m^{2}}\right)\!\right)\label{eq:i3}\\
\int_{\Lambda\mathrm{\mathrm{reg}}}\frac{d^{4}k}{i(2\pi)^{4}}\frac{k^{2}k_{\mu}k_{\nu}}{\left(k^{2}-m^{2}\right)^{3}} & \!\!=\!\! & -\frac{1}{(4\pi)^{2}}\frac{g_{\mu\nu}}{4}\left(2\Lambda^{2}-3m^{2}\ln\left(\frac{\Lambda^{2}+m^{2}}{m^{2}}\right)+m^{2}-\frac{m^{4}}{\Lambda^{2}+m^{2}}\right)\label{eq:i3a}\\
\int_{\Lambda\mathrm{\mathrm{reg}}}\frac{d^{4}k}{i(2\pi)^{4}}\frac{1}{\left(k^{2}-m^{2}\right)^{2}} & \!\!=\!\! & \frac{1}{(4\pi)^{2}}\left(\ln\left(\frac{\Lambda^{2}+m^{2}}{m^{2}}\right)+\frac{m^{2}}{\Lambda^{2}+m^{2}}-1\right)\label{eq:i4}\\
\int_{\Lambda\mathrm{\mathrm{reg}}}\frac{d^{4}k}{i(2\pi)^{4}}\frac{k_{\mu}k_{\nu}}{\left(k^{2}-m^{2}\right)^{3}} & \!\!=\!\! & \frac{1}{(4\pi)^{2}}\frac{g_{\mu\nu}}{4}\left(\ln\left(\frac{\Lambda^{2}+m^{2}}{m^{2}}\right)+\frac{m^{2}}{\Lambda^{2}+m^{2}}-1\right)\label{eq:i5}\\
\int_{\Lambda\mathrm{\mathrm{reg}}}\frac{d^{4}k}{i(2\pi)^{4}}\frac{k^{2}k_{\mu}k_{\nu}}{\left(k^{2}-m^{2}\right)^{4}} & \!\!=\!\! & \frac{1}{(4\pi)^{2}}\frac{g_{\mu\nu}}{12}\left(\!3\ln\left(\frac{\Lambda^{2}+m^{2}}{m^{2}}\right)\!+\!5\frac{m^{2}}{\Lambda^{2}+m^{2}}\!-\!\frac{m^{4}}{\left(\Lambda^{2}+m^{2}\right)^{2}}\!-\!4\!\right)\label{eq:i5a}\\
\int_{\Lambda\mathrm{\mathrm{reg}}}\frac{d^{4}k}{i(2\pi)^{4}}\frac{k_{\mu}k_{\nu}k_{\rho}k_{\sigma}}{\left(k^{2}-m^{2}\right)^{4}} & \!\!=\!\! & \frac{1}{(4\pi)^{2}}\frac{g_{\mu\nu}g_{\rho\sigma}\!+\! g_{\mu\rho}g_{\nu\sigma}\!+\! g_{\mu\sigma}g_{\nu\rho}}{24}\left(\!\!\ln\left(\frac{\Lambda^{2}\!+\! m^{2}}{m^{2}}\right)\!+\!\frac{m^{2}}{\Lambda^{2}\!+\! m^{2}}\!-\!1\!\right)\phantom{{12\}}}\label{eq:i6}\end{eqnarray}
(\ref{eq:i1}-\ref{eq:i3}) depend on the same function of $\Lambda$.
(\ref{eq:i2}, \ref{eq:i3}) are traced back to \eqref{eq:i1} via
\eqref{eq:rule1} and \eqref{eq:rule2}. \eqref{eq:i3a} and \eqref{eq:i5a}
have a different $\Lambda$ dependence. Evaluating these integrals
at first step \eqref{eq:rule0} is used, then \eqref{eq:rule1} or
\eqref{eq:rule2} can be applied to the remaining free indices.

\end{document}